\documentclass{aa}
\usepackage[utf8]{inputenc}
\usepackage[T1]{fontenc}

\usepackage[varg]{txfonts}
\usepackage{bbm}
\usepackage{graphicx}
\usepackage{if then}
\usepackage{longtable}
\usepackage{color}
\usepackage{tikz}
\usetikzlibrary{shapes,arrows}
\usepackage{verbatim}
\usepackage{tikz}
\usepackage{tikzscale}
\usepackage{filecontents}
\usepackage{algorithm,algorithmicx,algpseudocode}

\usepackage{todonotes}

\def\del#1{{}}

\newcommand{\ares}{{\sc\tt ARES}}

\definecolor{airforceblue}{rgb}{0.36, 0.54, 0.66}
 \definecolor{azure}{rgb}{0.0, 0.5, 1.0}

\begin{document}

\bibpunct{(}{)}{;}{a}{}{,} 

\title{Bayesian power-spectrum inference with foreground and target contamination treatment}
\author{ J. Jasche\inst{1} \and G. Lavaux\inst{2,3}}
\institute{ Excellence Cluster Universe, Technische Universit\"at M\"unchen, Boltzmannstrasse 2, 85748 Garching, Germany
\and CNRS \& Sorbonne Universit\'{e}s, UPMC Univ Paris 06, UMR7095, Institut d'Astrophysique de Paris, F-75014, Paris, France 
\and Sorbonne Universit\'es, Institut Lagrange de Paris (ILP), 98 bis bd Arago, 75014 Paris, France
}

\date{ Accepted 24/05/2017}

\label{firstpage}

\abstract{This work presents a joint and self-consistent Bayesian treatment of various foreground and target contaminations when inferring cosmological power-spectra and three dimensional density fields from galaxy redshift surveys. This is achieved by introducing additional block sampling procedures for unknown coefficients of foreground and target contamination templates to the previously presented \ares{} framework for Bayesian large scale structure analyses. As a result the method infers jointly and fully self-consistently three dimensional density fields, cosmological power-spectra, luminosity dependent galaxy biases, noise levels of respective galaxy distributions and coefficients for a set of a priori specified foreground templates. In addition this fully Bayesian approach permits detailed quantification of correlated uncertainties amongst all inferred quantities and correctly marginalizes over observational systematic effects. We demonstrate the validity and efficiency of our approach in obtaining unbiased estimates of power-spectra via applications to realistic mock galaxy observations subject to stellar contamination and dust extinction. While simultaneously accounting for galaxy biases and unknown noise levels our method reliably and robustly infers three dimensional density fields and corresponding cosmological power-spectra from deep galaxy surveys. Further our approach correctly accounts for joint and correlated uncertainties between unknown coefficients of foreground templates and the amplitudes of the power-spectrum. An effect amounting up to $10$ percent correlations and anti-correlations across large ranges in Fourier space.
}


\maketitle

\section{Introduction}
In recent years the cosmological community has witnessed great improvements in our understanding of the Universe. This progress is particularly due to the spectacular results of the Planck satellite mission and deep galaxy observations such as the ones provided by the Baryon Oscillation Sky Survey \citep[][]{PlanckMission,2013AJ....145...10D,Planck2015}.
These results put high standards for future analyses of cosmological data with an ever increasing need to control uncertainties and systematic effects in observations in order not to misinterpret data when searching for cosmological signals. To address these needs, data science is challenged to provide ever more robust data models accounting for complex systematic effects and allowing for accurate marginalization over unknowns when interpreting cosmological data.

A particular challenge in existing and coming deep galaxy redshift surveys arises from the need to properly understand selection processes of galaxies from which cosmological surveys are constructed \citep[see e.g.][]{2013MNRAS.432.2945H}. Such identification was conducted for mitigating star-galaxy contamination of the first SDSS photometric galaxy catalogue \citep{2002ApJ...579...48S}.
The problem is further exacerbated by our lack of understanding of galaxies as tracers of the underlying dark matter field when performing cosmological inference. 
In particular all our indicators for completeness rely on the relative slow, homogeneous and isotropic evolution of galaxy densities relative to dark matter densities. 
If the observation is further hindered by instrumental and/or terrestrial effect, this leads to a complex and challenging analysis problem. 

\begin{figure*}
	\begin{frame}
	    \centering
	    \resizebox{1.\textwidth}{!}{\includegraphics[]{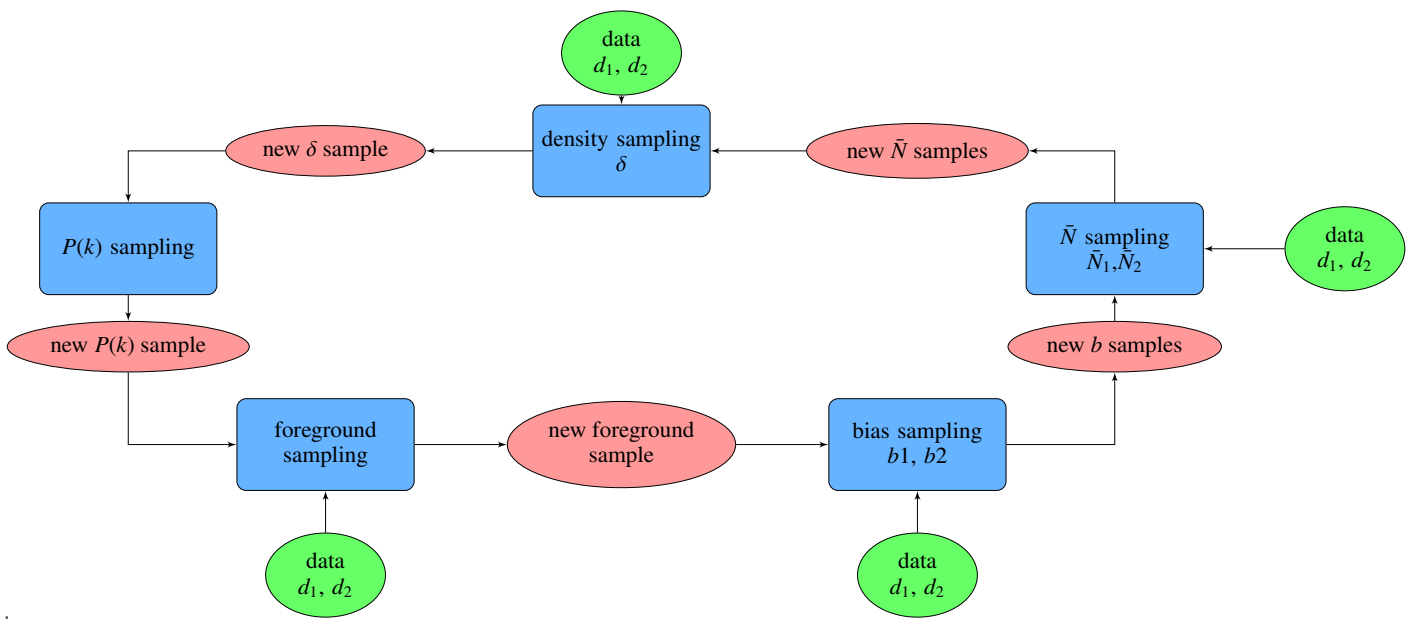}}
	\end{frame}
	\caption{Flow chart depicting the multi-step iterative block sampling procedure exemplified for 
    	two data sets. In the first step a three dimensional density field will be realized 
        conditional on the galaxy observations. In subsequent steps foreground template coefficients,
        the bias parameters, the power spectrum and the normalization parameters for the galaxy
        distribution are sampled conditional on respective previous samples. Iteration
		of this process yields samples from the full joint posterior distribution.	
        \label{fig:flowchart}
    }
\end{figure*}

In particular, \cite{Ross11} and \cite{Ho2012} have identified that contamination by bright stars alter significantly the intrinsic clustering signal of the observed photometric galaxy sample at large scales. The last SDSS release based on DR12 photometry still shows this problems in the measured correlation function \citep{BOSS_2017MNRAS.464.1168R}.
Effects due to foreground stars, dust, seeing, sky background intensity have the greatest potential to cause systematic deviations in the clustering signal \citep[see e.g.][]{2015JCAP...05..040H,2014MNRAS.444....2L}.
Selection of spectroscopic galaxies is not affected immediately by the same effect, but it is subject to other systematics, such as fibre collisions, target priority conflicts and fibre plate fixations.
All these data contaminations constitute a particular nuisance, since foreground affects are also affecting the noise properties of observed galaxy samples via varying attenuation or target contamination across the sky.

In the large scale structure community, foreground effects are traditionally treated by weighting observed galaxies to homogenize the distribution of the traced density field across the sky \citep[see e.g.][]{Ross11,2013MNRAS.433.1202S,BOSS_2017MNRAS.464.1168R}. This approach neglects possible and sometimes counter intuitive correlations between contamination effects and power-spectrum amplitudes across large ranges in Fourier space. It also ignores the effects of modified  observational noise properties due to target contamination. There have also been several methods proposed to account for additive contributions from unknown foregrounds in photometric and spectroscopic galaxy observations \citep[see e.g. ][]{1998ApJ...499..555T,2014MNRAS.444....2L,2015JCAP...05..040H}.
Also the literature on cosmic microwave background analyses provides a plenitude of approaches to account for linear additive foreground contributions \citep[see e.g.][]{1996MNRAS.281.1297T,2007ApJS..170..288H,2008ApJ...676...10E,2016arXiv161203401S,2016A&A...588A.113V,2017MNRAS.465.1847E}.

However, all these approaches do not properly account for multiplicative  foreground and target contaminations which also affect the noise. 
In this work we expand on the idea that foreground effects are more closely related to multiplicative rather than additive contributions. This is similar to the discussion presented in \citet{2013MNRAS.432.2945H}, who used a multiplicative correction in observed galaxy densities. 

In this work we seek to account for these effects when inferring cosmological power-spectra from observations. Literature provides a plenitude of various statistically more or less rigorous approaches to measure power-spectra. Several of these methods rely on Fourier transform based methods or exploit Karhunen-Loève or spherical harmonics decompositions \citep[see e.g.][]%
{FELDMAN1994,TEGMARK1995,HAMILTON1997A,YAMAMOTO2003,%
 PERCIVAL2004,TEGMARK1997,TEGMARK_2004,POPE2004,FISHER1994,HEAVENS1995,%
 TADROS1999,PERCIVAL2004,PERCIVAL2005%
 }. Other approaches aim at inferring the real space power spectrum via likelihood methods \citep[][]{BALLINGER1995,HAMILTON1997A,HAMILTON1997B,TADROS1999,PERCIVAL2005}.

In the Bayesian community several approaches have been proposed to jointly infer three dimensional density fields and their corresponding cosmological power-spectra \citep[see e.g. ][]{JASCHESPEC2010,2012MNRAS.421..251G,2016MNRAS.455.4452A}. Also note, that similar approaches explored for analyses of cosmic microwave background data \citep[see e.g.][]{WANDELT2004,ODWYER2004,2004ApJS..155..227E,JEWELL2004,LARSON2007,ERIKSEN2007,JEWELL2009}.

To account for such effects of foreground and target contamination in a statistically rigorous fashion, we propose a hierarchical Bayesian approach to jointly and self-consistently infer three dimensional density fields, corresponding power-spectra and coefficients of a set of different foreground templates. In particular this work builds upon our previously developed Bayesian inference algorithm \ares{} (Algorithm for REconstruction and Sampling) \citep[see e.g.][]{JASCHESPEC2010,JaschePspec2013,2015MNRAS.447.1204J,2016MNRAS.455.3169L}.

The manuscript is structured as follows. In Section~\ref{sec:statmodel} we give a brief overview of the statistical model that we propose. First we remind in Section~\ref{sec:ares} the hierarchical Bayesian inference approach on which our code, ARES, is based. Then we describe in Section~\ref{sec:contamination_model} the necessary modifications of the model for foreground effects and in Section~\ref{sec:fg_sampler} the modification to the original inference algorithm. In Section \ref{sec:mockdata} we describe the generation of artificial data used to test the performance of the sampling framework in section \ref{sec:results}. Finally we discuss our results and give an outlook on future applications in Section \ref{sec:conclusion}.

\begin{figure*}
	\includegraphics[width=\hsize]{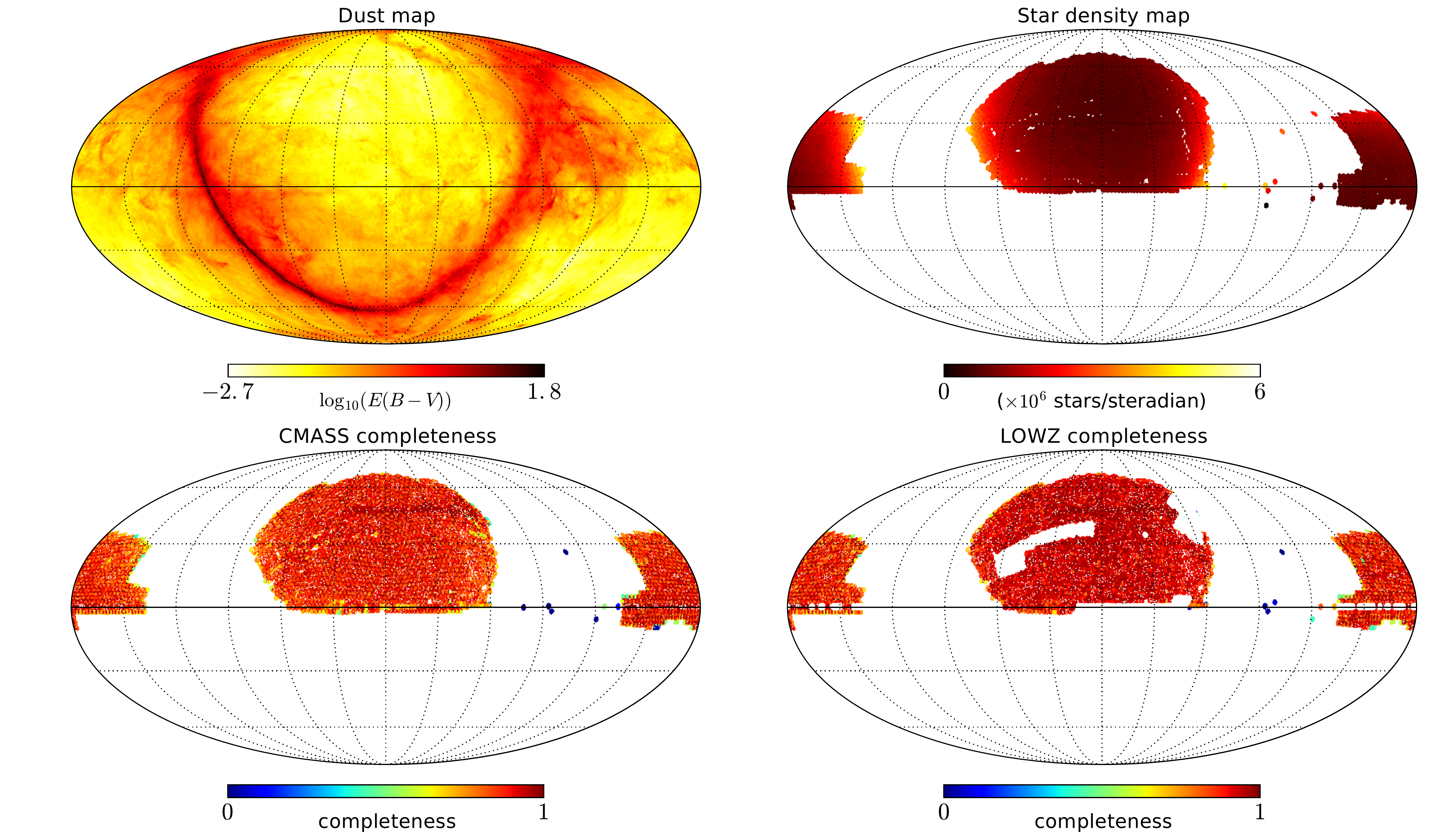}
	\caption{
    	We show here the foreground templates (top row) and the observed sky completenesses 
        (bottom row) used to generate and analyse the mock catalogue in this work. The upper 
        left panel shows the reddening map derived from the data of \protect\cite{SFD}. The upper 
        right panel is a star map count obtained as detailed in Section~\ref{sec:mockdata}. The 
        lower left panel gives the observed completeness for the mock CMASS survey and the lower right panel
        for the mock LOW-Z survey. These maps have been generated from SDSS-DR12 data \protect\citep{2011AJ....142...72E}. \label{fig:fig_msk_fg}
    }
\end{figure*}

\section{The Bayesian inference model}
\label{sec:statmodel}

This section provides a brief overview over our previously presented Bayesian inference framework \ares{} \citep[see e.g.][]{JASCHESPEC2010,JaschePspec2013,2015MNRAS.447.1204J}. We will also give a detailed description of the modifications enabling us to account for foreground and target contamination in deep galaxy observations.

\subsection{The \ares{} framework}
\label{sec:ares}
As discussed in the introduction, the current work builds upon the previously developed Algorithm
for Reconstruction and Sampling \ares{} \citep[see e.g.][]{JASCHESPEC2010,JaschePspec2013,2015MNRAS.447.1204J}. This full Bayesian large scale structure inference method aims at precision inference 
of cosmological power-spectra from galaxy redshift surveys. Specifically it performs joint inferences of three dimensional density fields, cosmological power spectra as well as luminosity dependent galaxy biases and corresponding noise levels for different galaxy populations in the survey \citep[][]{JASCHESPEC2010,JaschePspec2013,2016MNRAS.455.3169L}. 

In particular the \ares{} algorithm addresses a large scale data interpretation problem involving many millions of parameters.  In the following, for the sake of clarity, we will describe the corresponding data model in case of a single galaxy population. The generalization of this data model to account for an arbitrary number of galaxy populations with respective stochastic and systematic uncertainties has been described in our previous works  \citep[][]{JaschePspec2013,2016MNRAS.455.3169L}.
In case of a single galaxy population the data model is given as:
\begin{equation}
	N_i = \bar{N} R_i (1 + b D_i \delta_i) + \epsilon_i,
\end{equation}
where $N_i$ is the number of galaxies in the $i$-th grid element, $\bar{N}$ is the mean density of the galaxy population, $R_i$ is the overall linear response operator of the survey, describing redshift and target completeness, $b$ is the galaxy population bias, $D_i$ is the cosmic growth factor at the position of $i$-th grid element, $\delta_i$ is the density contrast at a reference redshift in this same grid element and $\epsilon_i$ denotes random but structured instrumental noise. Also, as described in our previous work, the observational noise will be assumed to be a Gaussian approximation to Poissonian noise, neglecting the influence of the signal itself. This assumption yields the corresponding noise covariance matrix as:
\begin{equation}
	\langle \epsilon_i \epsilon_j \rangle = \bar{N} R_i \delta^K_{i,j},
\end{equation}
with $\delta^K_{i,j}$ being the Kronecker-Delta (i.e. equal to one if $i=j$ and zero otherwise). Finally, we assume a homogeneous and isotropic Gaussian prior for density contrast amplitudes $\delta_i$. For further details please consult our previous work \cite{JaschePspec2013}.

In order to provide full Bayesian uncertainty quantification the algorithm explores the joint parameter space of density fields, power-spectra, galaxy biases and noise parameters via an efficient block sampling scheme.  
We show in Figure~\ref{fig:flowchart} a visualization of this iterative sampling procedure including the foreground sampling method presented in
this work. Iterative execution of these respective sampling steps provides us with
a valid Markov chain and a numerical representation of the full joint target posterior distribution. 
Also note, we use here an upgraded version of the \ares{} algorithm for which we employ the messenger method discussed in \citet{2013A&A...549A.111E}. This particular implementation of the Wiener posterior sampling method has been demonstrated to improve upon the statistical efficiency of previous implementations \citep[][]{2015MNRAS.447.1204J}.

\subsection{The foreground and target contamination model}
\label{sec:contamination_model}

Spectroscopic completeness is generally computed by the ratio of the number of observed spectra and the number of all photometric targets for a given area in the sky. This ratio is assumed to hold for any pointing of this area. However besides galaxies also a number of unknown contamination can contribute or affect observed photometric targets, artificially increasing or depleting their local number density. This contamination may include e.g. foreground stars, dust absorption or effects due to seeing. Naive estimate of the spectroscopic completeness from data therefore does not reflect the actual probability of obtaining a galaxy spectrum at a given position in the sky. From observations we can build an estimate of the completeness by calculating the ratio of the number of observed galaxy spectra $N^g_{i,\text{spectro}}$ and the number of \emph{target} galaxies $N^g_{i,\text{targets}}$, both in the direction of the pixel $i$ in the sky:
\begin{equation}
	C_{i,\text{obs}} = \frac{N^g_{i,\text{spectro}}}{N^g_{i,\text{targets}}} = \frac{N^g_{i,\text{spectro}}}{N^g_{i,\text{photo}}} \frac{N^g_{i,\text{photo}}}{N^g_{i,\text{targets}}} = C_i M^{-1}_i,
	\label{eq:contaminations}
\end{equation}
with $N^g_{i,\text{photo}}$ being the actual true number of galaxies that should have been identified by photometry and $M_i$ is the ratio of all the observed photometric targets $N^g_{i,\text{targets}}$ to the true sample of target galaxies $N^g_{i,\text{photo}}$. 
Equation~\eqref{eq:contaminations} demonstrates the dilemma, that given some spectroscopic information (for which objects can clearly be identified as galaxies) there is no immediate way to decide how many galaxies were in the actual target sample. In  Equation~\eqref{eq:contaminations} this mismatch is quantified by the ratio $M_i = N^g_{i,\text{targets}}/N^g_{i,\text{photo}}$ between the number of real photometric targets that should have been chosen vs the actual but unknown number of all galaxy targets.

We expect two possible contributions to $M_i$:
either there is an excess in the number of targets because the photometric information was insufficient to separate galaxies from stars or other objects in the sky, or there is a lack of galaxies that have not been detected due to, e.g., low surface brightness or dust absorption. Assuming such foreground contributions to be mild perturbations, the corresponding contamination map $M$ on the sky can be expressed as a product of small contaminants (greater or lesser than one). Individual contaminations can then be modelled via respective foreground templates. Note that this approach has also been adopted by the BOSS collaboration to correct their measurement of the cosmological power-spectrum \citep{Ross11,Ho2012,Anderson12}. Given this assumption we can express individual $M_i$ as:
\begin{equation}
	M_i = \prod_{n=0}^{N^\text{fg}} (1 - \alpha_n F_{n,i}),
	\label{eq:foreground_mod}
\end{equation}
where $F_{n,i}$ is the foreground template of the $n$-th contribution at the $i$-th pixel of the map, $\alpha_n$ is the amplitude of respective foreground templates, $N^\text{fg}$ is the total number of foreground maps. 
We note that different surveys or even different sub-samples of observed galaxies may be subjected to different foreground effects.
To consistently account for all these foreground effects when jointly analysing individual or several data sets the original data model implemented in \ares{} needs to be modified by a multiplicative correction of the survey response operator $R^c_i$.
Specifically, we model the observed number of galaxies $N^c_i$ in a survey as:
\begin{equation}
	N^c_i = \bar{N}^c M_i(\{\alpha^c_n\}) R^c_i (1 + b^c D_i \delta_i) + \epsilon^c_i\, ,
    \label{eq:full_data_model}
\end{equation}
where the additive noise contribution $\epsilon^c_i$ drawn from a zero mean Gaussian distribution with covariance matrix given as: 
\begin{equation}
	\langle \epsilon^c_i \epsilon^{c'}_j \rangle = \bar{N}^c M_i(\{\alpha^c_n\}) R^c_i \delta^K_{i,j} \delta^K_{c,c'}.
\end{equation}
The superscript $c$ indicates different considered catalogues.
These modifications render the posterior distribution of $\delta_i$ more complex. By construction the likelihood is given as:
\begin{multline}
	\mathcal{L}(\{N^c_i\} | \{\delta_i\}, \{\bar{N}^c\},\{b^c\}) \propto 
	 \prod_{c=0}^{N^c} \prod_{i=0}^{N^v} \left( \bar{N}^c M_i(\{\alpha^c_n\}) R^c_i \right)^{-1/2} \\
	\exp\left\{-\frac{1}{2} \frac{1}{\bar{N}^c M_i(\{\alpha^c_n\}) R^c_i} \right. \\
	\left. \left[N^i_c -  \bar{N}^c M_i(\{\alpha^c_n\}) R^c_i (1 + b^c D_i \delta_i) \right]^2 \right\}.
	\label{eq:foreground_LH}
\end{multline}
It should be remarked that foreground contributions, as modelled here, are not just mere additive contributions to the signal to infer, but they also have pronounced impact on the varying noise properties across the survey. Hence, as can be seen from Equation~\eqref{eq:foreground_LH} inferring the foreground coefficients $\{\alpha_n\}$ is a highly non-linear analysis task. 

\begin{figure}
 {\centering
  \includegraphics[width=\hsize]{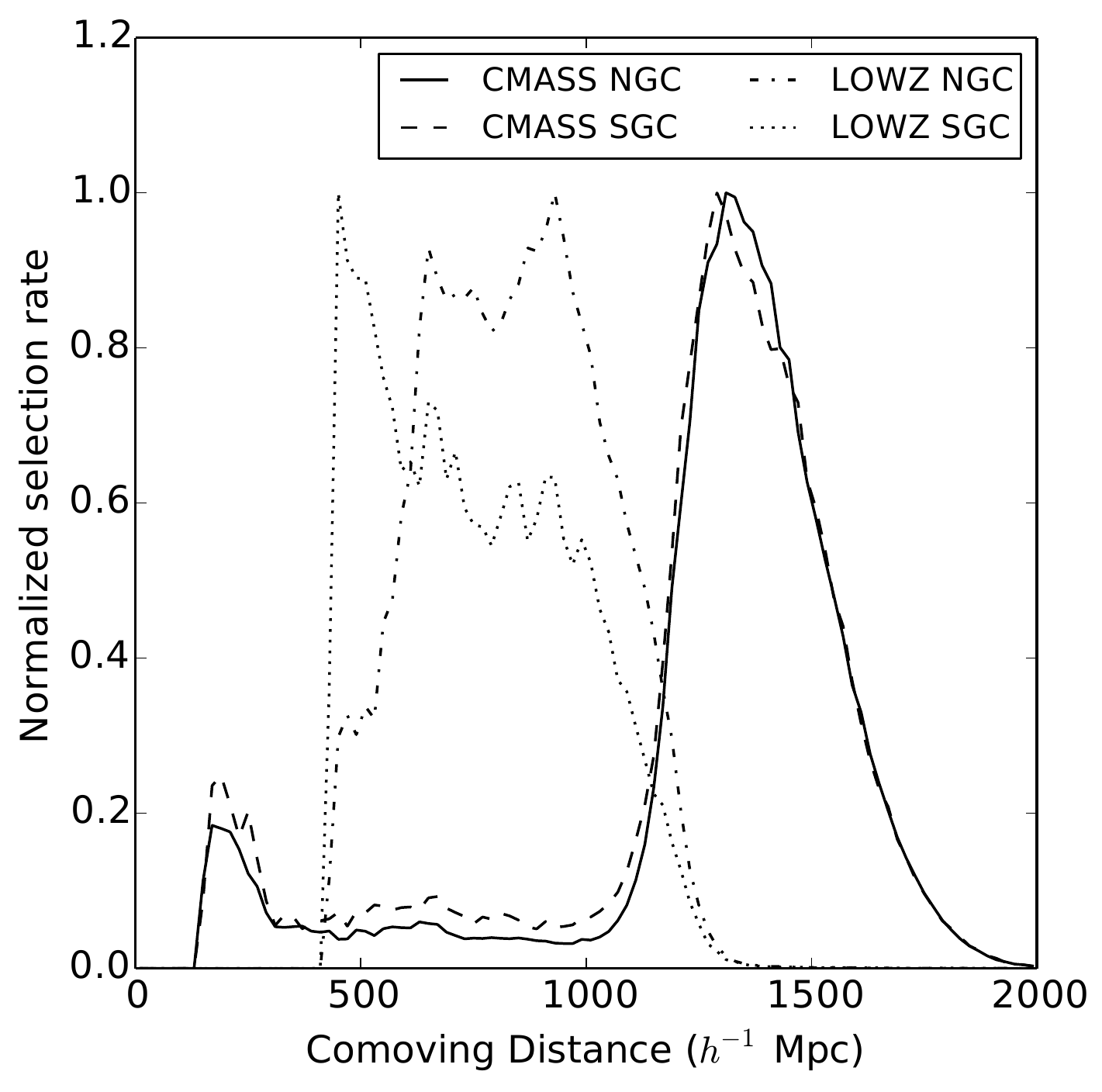} 
 }
  \caption{The plot gives the radial selection functions for the CMASS and LOW-Z sample that we have used to generate the mock data to closely resemble the actual SDSS3 BOSS data. The CMASS selection is given in thick solid (North Galactic Cap) and dashed line (South Galactic Cap). The LOWZ selection is given in thin dash-dotted (North Galactic Cap) and dotted lines (South Galactic Cap). \label{fig:radial_selfunc}}
\end{figure}

\subsection{Sampling foreground coefficients}
\label{sec:fg_sampler}

As described by the likelihood distribution given in Equation~\eqref{eq:foreground_LH} foregrounds of different catalogues as labelled by the superscript $c$ can be sampled independently. For this reason and without loss of generality here we provide the sampling procedure for a single galaxy catalogue only. Then the conditional posterior distribution for the coefficients $\{\alpha_n\}$ of respective foreground templates can be written as:
\begin{multline}
\mathcal{P}\left( \{\alpha_n\}| \{N_i\}, \{\delta_i\}, \{\bar{N}\},\{b\}\right) \propto \\ 
  \mathcal{P}\left( \{\alpha_n\}\right ) \, \mathcal{L}(\{N_i\} | \{\delta_i\}, \{\bar{N}\},\{b\},\{\alpha_n\}) \, ,
\label{eq:foreground_LH_single}
\end{multline}
where $ \mathcal{P}\left( \{\alpha_n\}\right ) $ is the prior of foreground coefficients and $\mathcal{L}(\{N_i\} | \{\delta_i\}, \{\bar{N}\},\{b\},\{\alpha_n\}) $ is the likelihood for a single catalogue as given by Equation~\eqref{eq:foreground_LH}.
In the absence of any further information on the amplitudes of foreground coefficients $\alpha_n$ we follow the maximal agnostic approach by using a uniform prior $ \mathcal{P}\left( \{\alpha_n\}\right )=1$. 
It can be seen from Equations~\eqref{eq:foreground_LH_single} and \eqref{eq:foreground_LH} that conditional posterior distribution does not factorize in the coefficients $\alpha_n$ for respective foreground templates. 
To correctly account for the conditional dependencies between the coefficients $\{\alpha_n\}$, we propose to use a block sampling procedure to sequentially draw random variates of $\alpha_n$ with respect to all other values.
 This is achieved by introducing the following sequence of sequential sampling steps to the full \ares{} framework:
 \begin{align}
 \alpha_0 & \sim\mathcal{P}\left(\alpha_0| \{\alpha_n\}\setminus \alpha_0, \{N_i\}, \{\delta_i\}, \{\bar{N}\},\{b\}\right) \nonumber \\
 \alpha_1 & \sim\mathcal{P}\left(\alpha_1| \{\alpha_n\}\setminus \alpha_1, \{N_i\}, \{\delta_i\}, \{\bar{N}\},\{b\}\right)\nonumber \\
 \alpha_2 & \sim\mathcal{P}\left(\alpha_2| \{\alpha_n\}\setminus \alpha_2, \{N_i\}, \{\delta_i\}, \{\bar{N}\},\{b\}\right)\nonumber \\
& \vdots \nonumber \\
 \alpha_{N^{fg}-1} & \sim \mathcal{P}\left(\alpha_{N^{fg}-1}| \{\alpha_{n}\}\setminus \alpha_{N^{fg}-1}, \{N_i\}, \{\delta_i\}, \{\bar{N}\},\{b\}\right) \nonumber
 \end{align}
 where the symbol "$\sim$" indicates a random draw from respective distributions. This sampling procedure integrates well into the \ares{} framework as indicated in Fig.~\ref{fig:flowchart}.
Despite the fact that drawing respective realizations of the foreground coefficients $\{\alpha_n\}$ is a non-linear process, there exists a direct sampling procedure. The detailed derivation of the foreground coefficient sampler is presented in Appendix~\ref{app:fg_sampler}. The detailed algorithm for generating respective random variates is given in Algorithm~\ref{cond_fg_sampler}.

\section{Generation of Gaussian mock data}
\label{sec:mockdata}

To test the validity and performance of the modified \ares{} sampling framework we follow a similar approach as discussed in our previous works \citep[][]{JASCHESPEC2010,JaschePspec2013,2015MNRAS.447.1204J}. 
These mock catalogues are generated in accordance with the data model described in Equation~\eqref{eq:foreground_LH}, including various foreground effects.
We generate artificial galaxy data on a cubic equidistant grid of side length \(4000 \,h^{-1}\mathrm{Mpc}\) consisting of \(256^3\) grid nodes.
 
First a realization of a cosmic density contrast field \(\delta_i\) is drawn from a zero-mean normal distribution with covariance matrix corresponding to a cosmological power-spectrum. This spectrum, including baryon acoustic oscillations, is calculated according to the prescription described in \citet{1998ApJ...496..605E} and \citet{1999ApJ...511....5E}. For numerical evaluation we  assume a \(\Lambda\)CDM cosmology  with the set of parameters given as  (\(\Omega_m=0.3089\), \(\Omega_{\Lambda}=0.6911\), \(\Omega_{b}=0.0485\), \(h=0.6774\), \(\sigma_8=0.8159\), \(n_s=0.9667 \)), as determined by Cosmic Microwave Background observations of the Planck satellite mission  \citep{Planck2015}.

\begin{figure}[t]
	\includegraphics[width=\hsize]{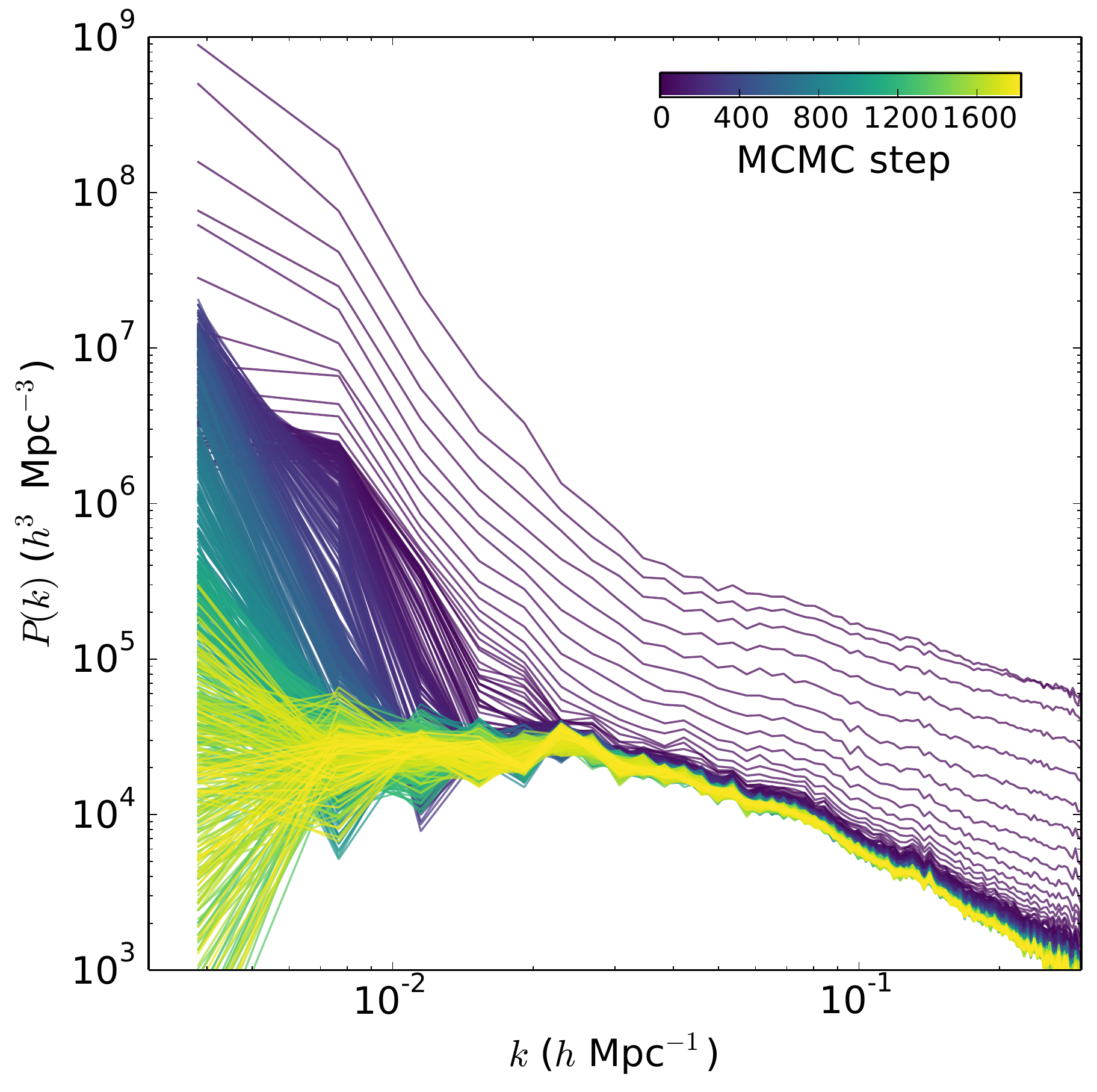}
	\caption{The plot shows the sequence of posterior power-spectrum realizations during the initial sampling steps of the Markov chain as indicated by the colour-bar in the plot. As can be seen the Markov chain performs a coherent drift from an over-disperse initial state towards the preferred region in parameter space. This initial burn-in phase lasts for $\sim 2000$ Markov transitions. \label{fig:burnin_spec_corrlength}}
\end{figure}

Following a similar description as discussed in \citet{JaschePspec2013},  we intent to create a realistic scenario to jointly analyse
the SDSS DR7 main, the CMASS and the LOW-Z galaxy sample while accounting for their respective systematic effects.  Corresponding artificial data sets are then drawn from the distribution given in Equation~\eqref{eq:foreground_LH}. These artificial data sets include the effects of noise, galaxy bias, survey geometries, selection and foreground effects. For the sake of this work we restrict our tests to accounting for the dominant foreground effects of dust extinction and stellar contamination. The templates for these two effects are presented in the upper panels of Fig. \ref{fig:fig_msk_fg} the lower panels show the completeness masks for the respective surveys. 

\begin{figure*}
	\includegraphics[width=\hsize]{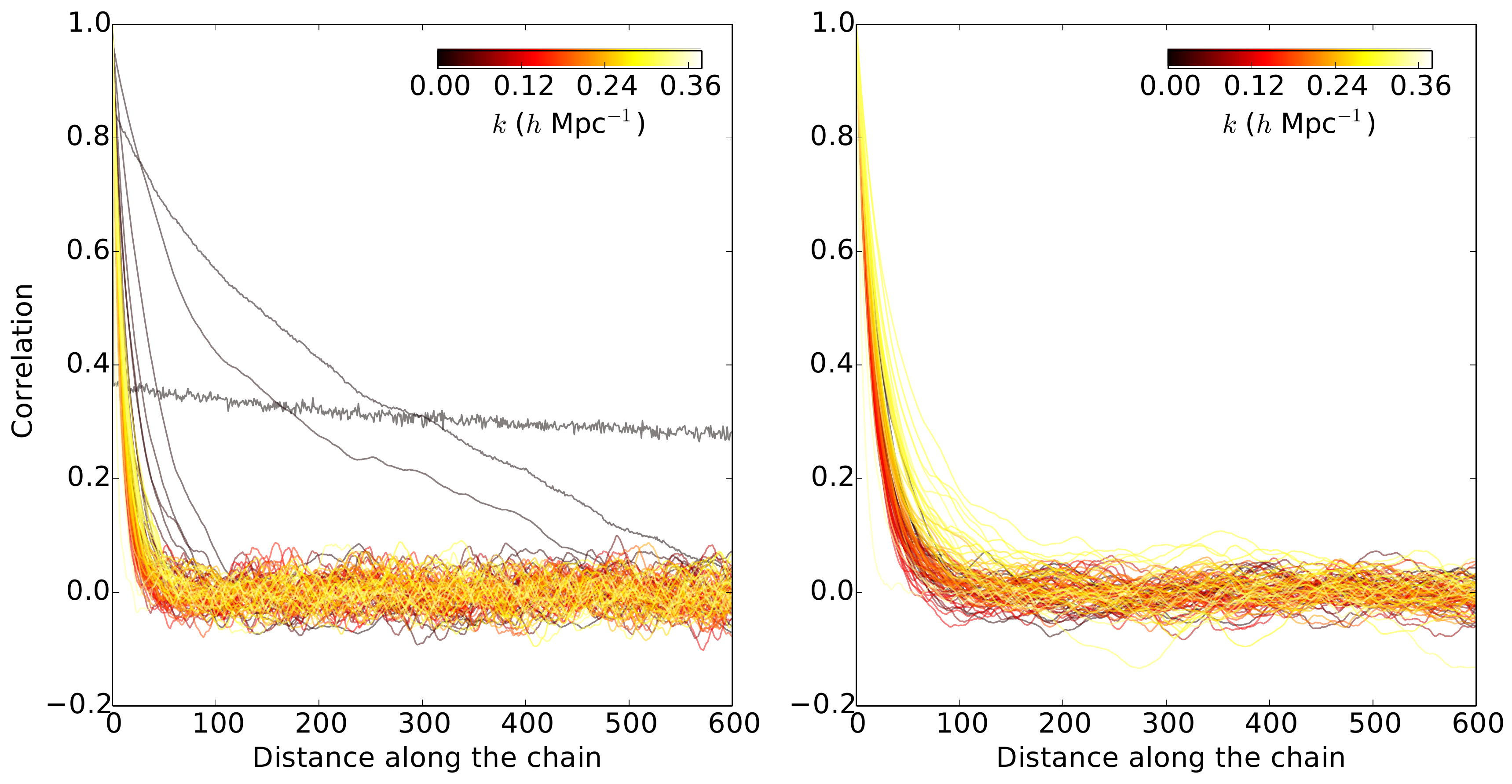}
	\caption{Correlation length of the sampler for different modes as indicated by the colour bar on the left. As can be seen most of the modes have typical correlation lengths on the order of $xxx$ samples. Some large scale modes exhibit longer correlation length as explained in the text.\label{fig:corrlength}}
\end{figure*}

The map describing dust extinction has been generated straightforwardly from the SFD maps \citep{SFD}. In particular we constructed a HEALPix map of the reddening at $N_\text{side}=2048$ by linearly interpolating the values of the SFD map \citep{HEALPIX}. The star map is built from different pieces of information. The first component consists in computing a MANGLE description\footnote{MANGLE software is originally provided by \citet{SWANSON2008MNRAS}.} of the geometry of the spectroscopic plates. We then count the number of stars with apparent magnitudes $20.3<i_\text{PSF}<20.6$ present in each single non-overlapping polygon. We convert these MANGLE description into an HEALPix map and divide the value in each pixel by the area of the overlapping polygon. This results in a map for which each pixel has an estimated star count by steradians, thus a star density. We have chosen to average by spectroscopic plates to reduce shot-noise in the estimate. A better estimate would have been obtained from the geometrical description of the photometric tiling but at the cost of increased noise.

Further, for the SDSS DR7 main sample component of the mock data we assume a radial selection function following from a standard Schechter luminosity function with standard r-band parameters ($\alpha = -1.05$, $M_* -5 \mathrm{log}_{10}(h)=-20.44$), and we limit the survey to only include galaxies within an apparent Petrosian r-band magnitude range  $13.5 < r < 17.6$ and within the absolute magnitude ranges $M_{min}=-17.0$ to $M_{max}=-23.0$. As usual, the radial selection function $f(z)$ is then given by the integral of the Schechter luminosity function over the range in absolute magnitude.

For the CMASS and LOW-Z component we have used numerical estimates of the selection functions by computing a histogram of the corresponding $N(d)$ in the actual data sets \citep[e.g. for DR12][]{BOSS_2017MNRAS.464.1168R} ($d$ being the co-moving distance from the observer). To account for the different selection effects in the northern and southern galactic plane we also split our mock data sets into the CMASS and LOW-Z catalogues correspondingly. The respective radial selection functions are presented in Fig. \ref{fig:radial_selfunc}.  
The average of the product of the two dimensional survey geometry $C^c(\hat{x}) = C^c_{i(\hat{x})}$ and the selection function $f(x)$ at each grid element in the three dimensional volume yields the survey response operator:
\begin{equation}
  R^c_i = \frac{1}{|\mathcal{V}_i|} \int_{\mathcal{V}_i} \text{d}^3 \vec{x}\; C^c(\hat{x}) f^c(x) \,,
\end{equation}
with $|\mathcal{V}|$ the volume of the set $\mathcal{V}$, $\mathcal{V}_i$ indicating the volume represented by the $i$-th grid element. 

Given these definitions and a realization of the three dimensional density field \(\delta_i\), realizations of artificial galaxy observations for respective catalogues labelled with $c$ can be obtained by evaluating:
\begin{equation}
  N^c_i= \bar{N}^c\,R^c_i\,M^c_i\,(1+ b^c \delta_i) + \sqrt{\bar{N}^c\,R^c_i M^c_i}\, \epsilon_i \, , \label{eq:data_model}
\end{equation}
where $\epsilon_i$ is a white-noise field drawn from zero-mean and unit variance normal distribution.

\section{Results}
\label{sec:results}
In this section we discuss results obtained by applying the modified \ares{} algorithm to artificial mock data. In particular in this work we focus on the validity and statistical efficiency of the algorithm.

\subsection{Statistical efficiency of the sampler}
\label{sec:stat_eff}

\begin{figure*}
 \centering
  \includegraphics[width=1.\linewidth]{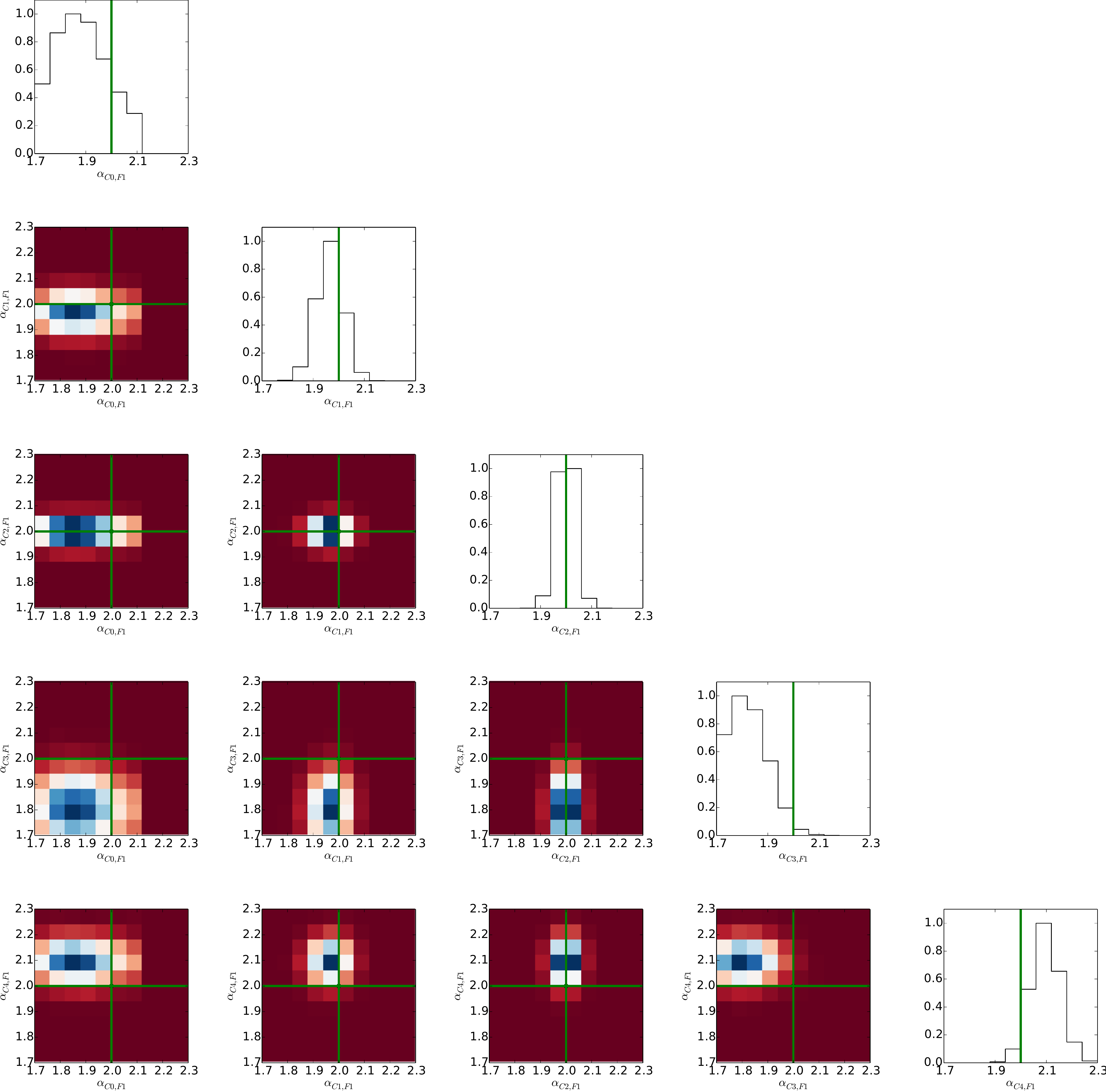}
  \caption{Marginalized 2d probability distributions for foreground contamination coefficients of the five mock catalogues used in this analysis. Green lines indicate the true values of foreground coefficients as used for the generation of the artificial mock galaxy survey. As can be seen different panels show various degrees of correlation between inferred coefficients that are correctly accounted for by our Markov chain. }
 \label{fig:marginalized2d_pdf}
\end{figure*}

To test the statistical efficiency of our sampler we follow a standard test procedure as described in previous works \citep[see e.g.][]{JASCHESPEC2010,JaschePspec2013,2015MNRAS.447.1204J}. In particular we test the initial burn-in phase by starting the sampler from an over dispersed state and monitoring transitions in parameter space as a sequence of steps in the Markov chain. Typically this test reveals a coordinated drift of inference parameters towards their target values. Once the chain moved to a preferred region in parameter space it starts to correctly explore the target posterior distribution via the random walk Gibbs sampling approach. At this stage it is assumed that the sampler has passed the initial burn-in phase and we start recording samples of the Markov chain. In Fig. \ref{fig:burnin_spec_corrlength} we show the sequence of sampled posterior power-spectra during the burn-in phase. For this test we started from an over dispersed state by multiplying the initial guess of a Gaussian random density contrast field by a factor of $0.1$. Fig. \ref{fig:burnin_spec_corrlength} nicely demonstrates the initial drift towards the preferred region in parameter space. As can be seen the initial burn in phase consists of about $\sim 2000$ Markov steps. As a side remark we note that generation of individual samples require an investment of $\sim 1.87$  CPU-hours per sample for the present scenario of dealing with five different galaxy sub-catalogues and two foregrounds.

To further test the statistical efficiency of the Gibbs sampling procedure we estimated its efficiency in generating independent Markov samples. Generally subsequent samples in a Markov chain are correlated and do not qualify for independent samples of the target posterior distribution. To estimate how many independent samples can be drawn from a Markov chain with a given number of transition steps one has to determine the length over which sequential samples are correlated. This correlation length characterizes the statistical efficiency of generating independent realizations of a parameter $\theta$ as follows:
\begin{equation}
\label{eq:CORR_COEFF}
C(\theta)_n =\frac{1}{N-n} \sum_{i=0}^{N-n}  \frac{\theta^i-\left \langle \theta\right \rangle}{\sqrt{\mathrm{Var} \left(\theta\right)}} \frac{\theta^{i+n}-\left \langle \theta\right \rangle}{\sqrt{\mathrm{Var} \left(\theta\right)}}  \, ,
\end{equation}
where \(n\) is the distance in the chain measured in iterations, $\left \langle \theta\right \rangle = 1/N \sum_i \theta^i$  and  $\mathrm{Var} \left(\theta\right) =  1/N \sum_i \left(\theta^i -\left \langle \theta\right \rangle \right)^2 $ and $N$ is the total number of samples in the chain.

As an illustration in Fig. \ref{fig:corrlength} we show the correlation length for the power-spectrum amplitudes of different modes in Fourier space, as indicated in the plot. It can be seen that the typical correlation length for a BOSS like survey analysis is on the order of $\sim 100$ Markov transitions. These results demonstrate the numerical feasibility of complex full Bayesian analyses of present and next generation surveys. 

\begin{figure*}
\centering
 \includegraphics[width=\hsize]{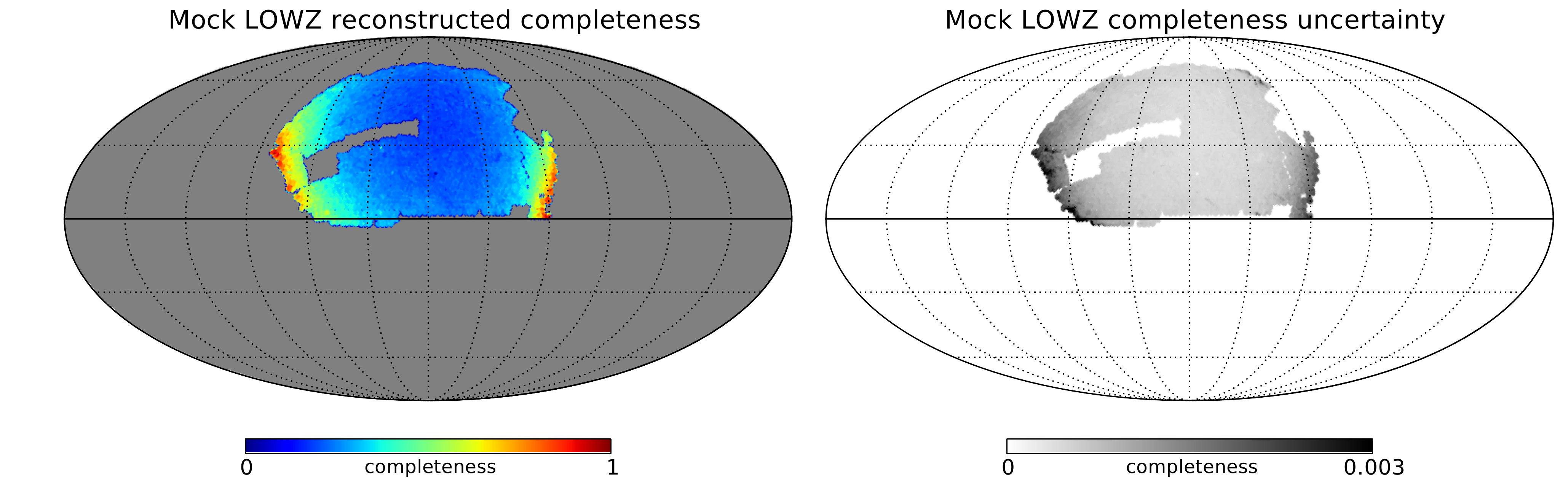}
  \caption{Ensemble mean of the effective survey response operator (left panel) and corresponding standard deviation map (right panel). The ensemble mean is renormalized by the largest pixel value, as the absolute value does not have an meaning independent of the mean density $\bar{N}$ and the radial selection function. The two above maps should be compared to the north galactic cap of the map in the lower right panel of Figure~\ref{fig:fig_msk_fg}. The ensemble mean is quite different ought to the introduced star contamination which could introduce contamination in targets. This manifests itself by an over-completeness on the edge of the map. The right map shows a similar trend but touching the uncertainty on the selection this time. \label{fig:mean_var_effective_survey response}
}
\end{figure*}

\begin{figure*}
\centering
  \includegraphics[width=\hsize]{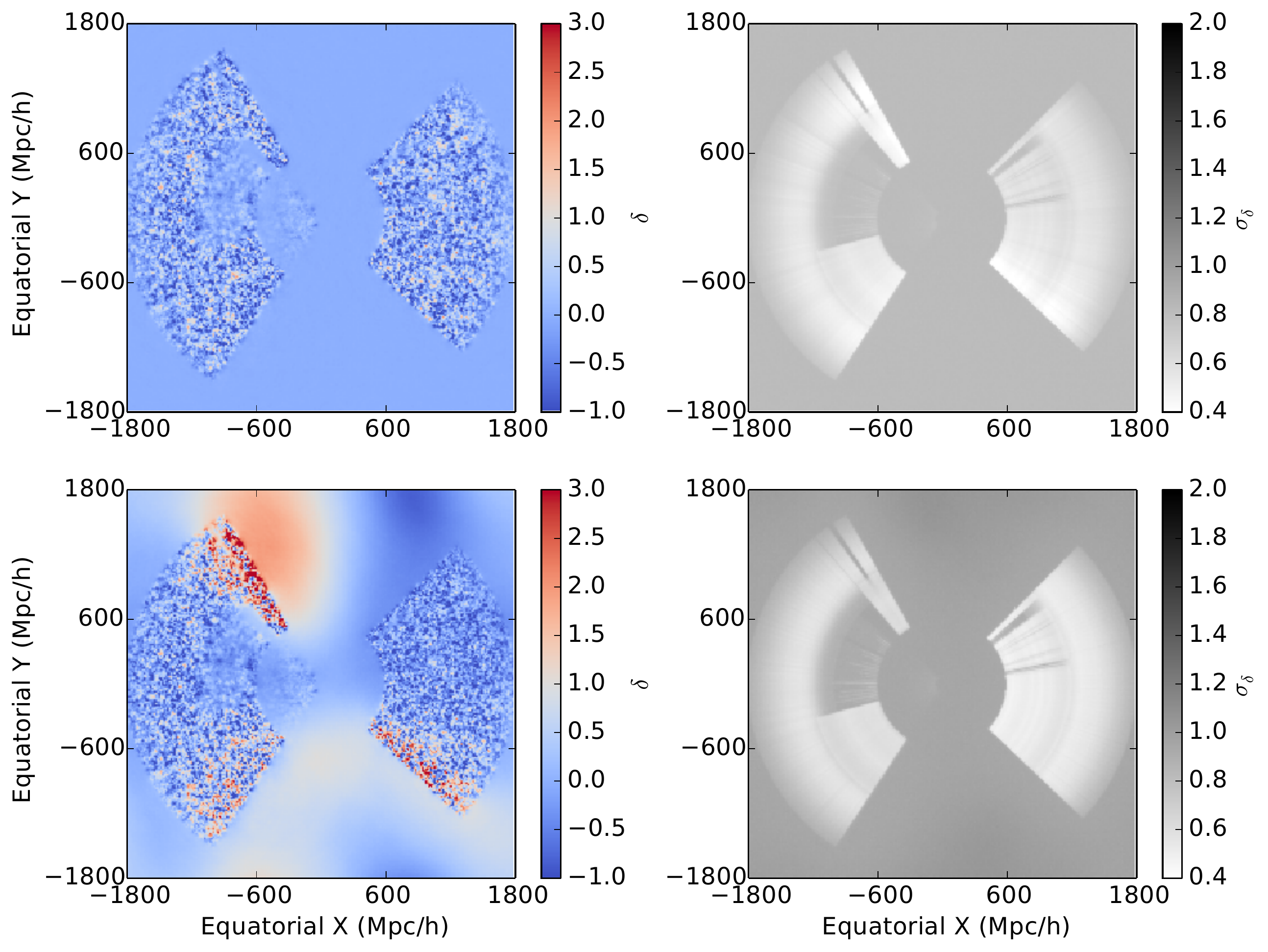}
  \caption{Slices through three dimensional ensemble mean (left panels) and variance fields (right panels). Top panels show results obtained with foreground correction while bottom panels show results without any foreground correction. As for the power-spectrum, we find an excessive large scale power when foreground corrections are not applied. Computing the foreground self consistently results in a non contaminated reconstruction. The variance fields are also affected as it is possible to see that the bottom is notably darker in average compared to the top slice, which indicates higher variance. }
  \label{fig:mean_var_dens}
\end{figure*}

\begin{figure*}
\centering
 \includegraphics[width=\hsize]{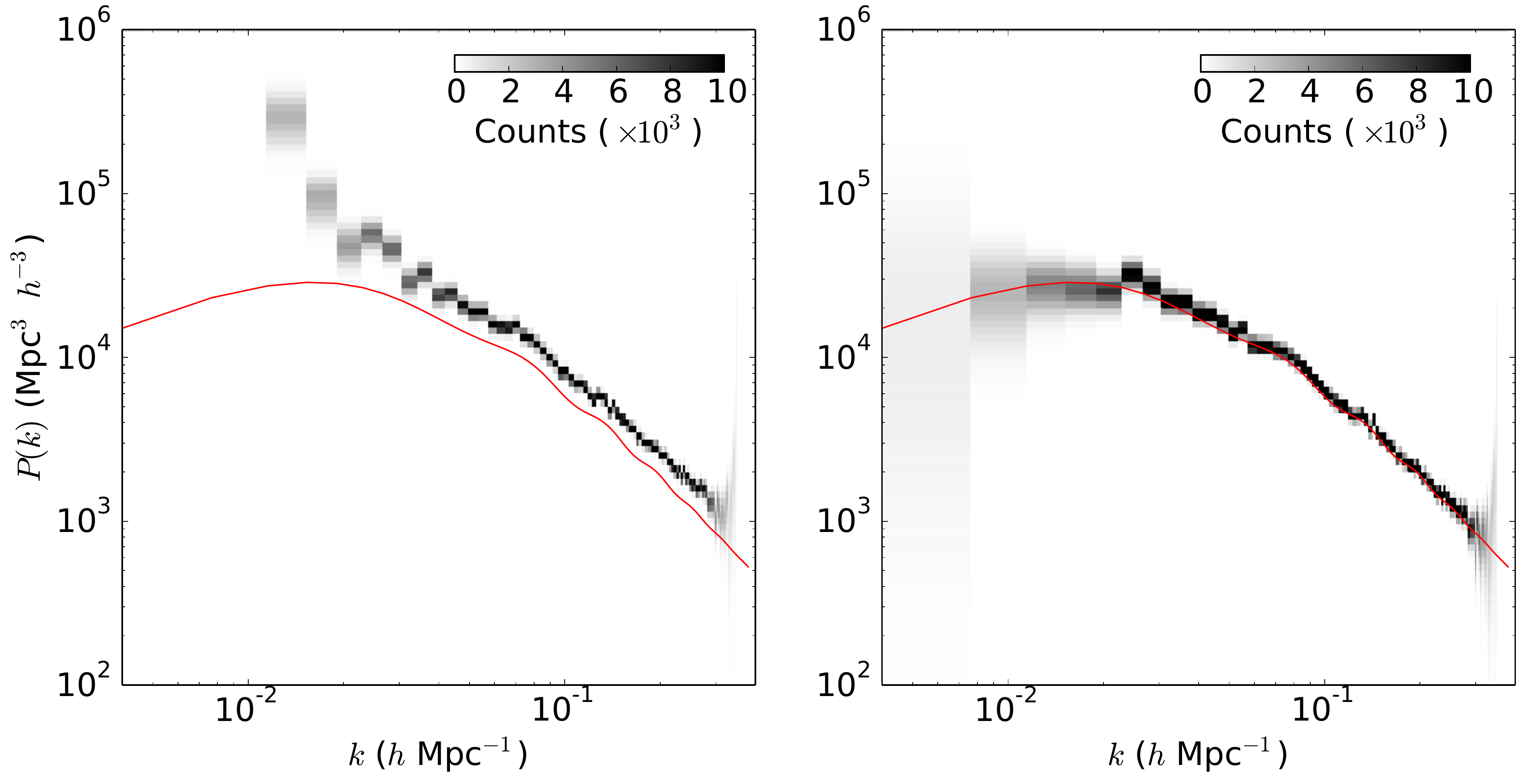}
  \caption{
  Univariate posterior distributions of power-spectrum amplitudes for a test without (left panel) and with (right panel) foreground corrections over the full range of Fourier modes considered in this work. Red lines correspond to the true underlying cosmological power-spectrum from which mock data sets were generated. The left panel clearly shows that uncorrected foreground effects yield excessive power for large scale modes and also introduces an overall biased result. In contrast the right panel shows results obtained from our test with foreground corrections. It can be clearly seen, that detailed treatment of all foreground effects permits us to obtain an unbiased measurement of power-spectrum amplitudes over the full range of Fourier modes.
  }
  \label{fig:posterior_spectra}
\end{figure*}

\subsection{Inference of template coefficients}
\label{sec:temp_coeffs}
As discussed above the proposed hierarchical Bayesian inference machine aims to account for the uncertainties of systematic effects arising from foreground effects. In particular the \ares{} framework correctly accounts for all joint and correlated uncertainties of different inference parameters and even across the five different mock galaxy surveys as used in this work. To illustrate this fact in Fig. \ref{fig:marginalized2d_pdf} we present two dimensional marginal posterior distributions for the corresponding foreground template coefficients. As can be seen, different panels indicate various degrees of correlation between different foreground coefficients across the five mock catalogues. We would also like to point out that generally, distributions for foreground template coefficients are highly non-Gaussian, and can have sharp transitions due to the requirement that effective contamination templates are required to have a positive sign.

As an interesting by-product of our sampling procedure we are able  to provide effective survey response maps which account for the a priori unknown systematics due to foreground and target contamination. In particular mean and standard deviations of such maps can be estimated by evaluating Equation~\eqref{eq:foreground_mod} for every foreground template parameter coefficient in the Markov chain and multiplying it with the estimated completeness map $C_{i,obs}$. The result is demonstrated in Fig.~\ref{fig:mean_var_effective_survey response}. It can be seen that most corrections appear at the boundary of the mask. These are the regions most affected by foreground stars in our test scenario. Note, that corresponding standard deviations, as shown in the right panel of Fig. \ref{fig:mean_var_effective_survey response}, also show increased uncertainty in these regions. This demonstrates that the algorithm accounts for larger uncertainty for more unreliable portions of the data, and optimally extracts cosmological information from observations, as discussed in the following.

\subsection{Inferred three dimensional density fields}
\label{sec:density_field}
Although this work focusses primarily on the inference of cosmological power-spectra it is instructive to also look at inferred three dimensional density fields. In particular we would like to highlight the impact of foreground contaminations on the inference of density fields from galaxy redshift surveys. To do so we compare two \ares{} analyses of the generated mock galaxy catalogue, with and without foreground treatment. From these two Markov chains we can calculate the ensemble mean density and corresponding variance fields. Results are presented in Fig. \ref{fig:mean_var_dens}. As can be seen, the analysis without a detailed treatment of foreground contaminations shows residual large scale features and erroneous power particularly close to the survey boundaries. These regions are affected the most by stellar contamination as indicated by the corresponding foreground map shown in Fig.~\ref{fig:fig_msk_fg}. 

In contrast our \ares{} run with detailed Bayesian foreground treatment shows a homogeneous density distribution throughout the entire observed domain.
Also note that variance maps of the corrected and uncorrected  \ares{} run look very similar. Apparently the erroneous features in the uncorrected \ares{} analysis do not affect the corresponding variance map. Since for the uncorrected run, the data model does not account for the systematic uncertainties associated to foreground contaminations, the reconstructed erroneous large scale power in the field will be fully attributed to the inferred large scale structure. From Fig. \ref{fig:mean_var_dens} it is visually evident that our data model including the treatment of foreground contaminations is much more robust against such misinterpretations. In the following we will have a closer look at inferred cosmological power-spectra.

\subsection{Inferred cosmological power-spectra}
\label{sec:power_spectra}
One of the most important features of the \ares{} algorithm is its ability to jointly infer three dimensional density fields, corresponding cosmological power-spectra, galaxy biases, noise levels and coefficients of several foreground templates including a detailed treatment of all joint and correlated uncertainties. Since the \ares{} framework yields proper Markov chains we are able to correctly marginalize over all joint uncertainties, when focussing on the analysis of specific target quantities such as the cosmological power-spectrum. 
Specifically in our previous work  \cite{JaschePspec2013} we have already demonstrated that the \ares{} algorithm reveals and correctly treats the anti-correlation between bias amplitudes and power spectrum, a 20\% effect across large ranges in Fourier space. Here we also take into account the unknown coefficients of several foreground templates.

To study the impact of foreground contamination on the analyses of cosmological power-spectra in deep galaxy surveys we compare inference results obtained from our two \ares{} runs with and without corresponding corrections. The results for inferred power-spectra are presented in Fig.~\ref{fig:posterior_spectra} where we show the univariate marginal posterior distribution for power-spectrum amplitudes at different modes in Fourier space. For the Markov chain without foreground treatment one can clearly observe excessive power at the largest scales. This observation corresponds to the excessive large scale power observed in corresponding inferred three dimensional density fields, as discussed above. In addition one can observe a slight bias with respect to the true underlying power-spectrum from which mock observations were generated. This can easily be understood by inspecting the data model described in Equation~\eqref{eq:full_data_model}, where one can see that there is a certain potential for a degeneracy between foreground coefficients and galaxy biases. If foreground effects are not treated correctly, then some of the foreground contributions will erroneously be compensated by sampled galaxy bias amplitudes, introducing the offset between true and recovered power-spectra shown in Fig. \ref{fig:posterior_spectra}. In contrast inferred power-spectra for the run with foreground treatment are unbiased with respect to the true underlying power-spectrum over the full domain of Fourier modes considered in this work. In particular the shape of the recovered power-spectrum at the largest scales is in excellent agreement with the true fiducial model.

\begin{figure}
\centering
  \includegraphics[width=1.\linewidth]{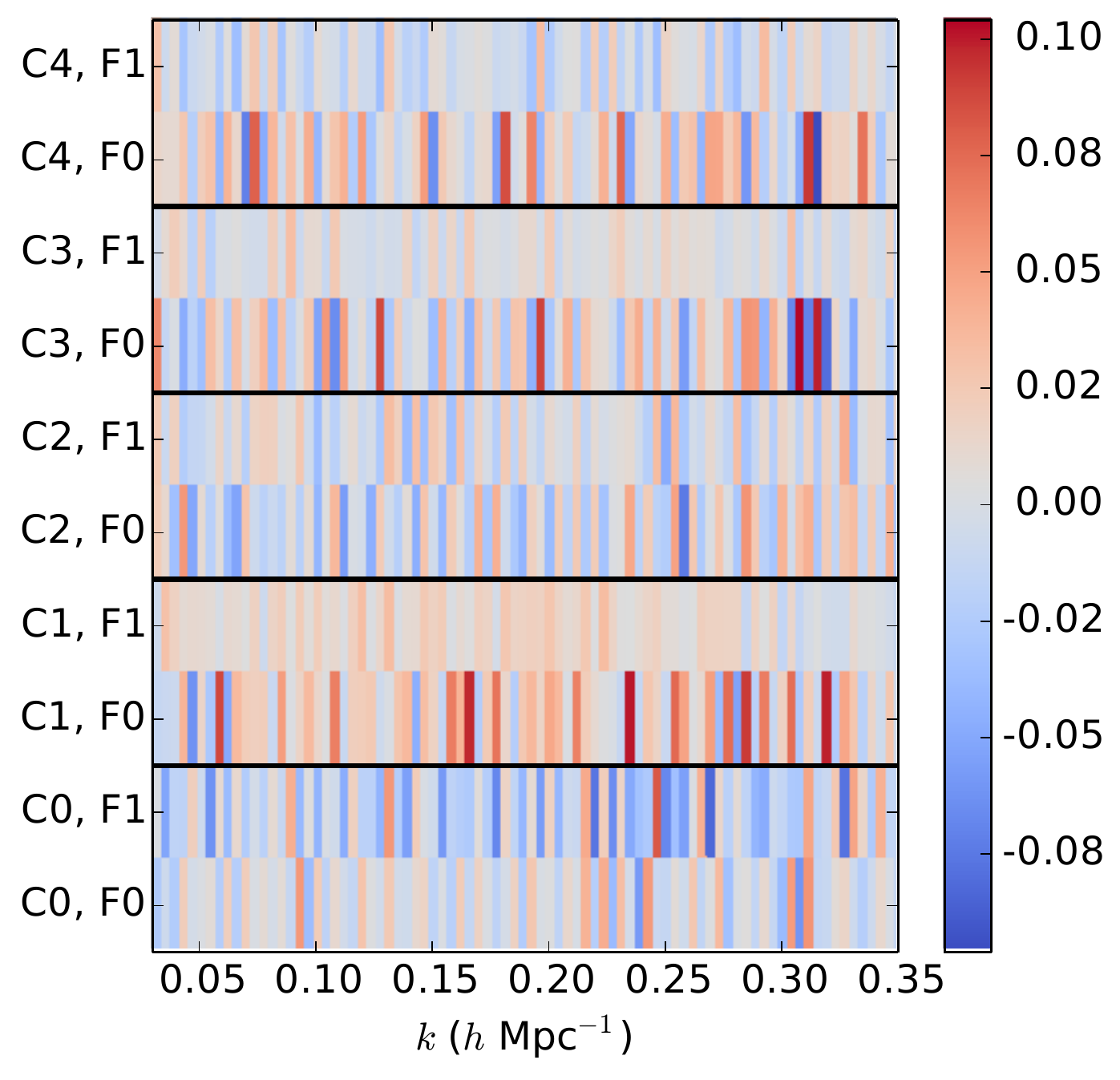}
  \caption{Cross correlation between power-spectrum amplitudes at different Fourier modes and negative (labelled F0) as well as positive  (labelled F1) foreground coefficients for the five respective catalogues (labelled C0 to C4).  \label{fig:corrmat_spec_fg}}
\end{figure}

We further studied the impact of different foreground coefficients on power-spectrum amplitudes at different scales in Fourier space. In particular we calculated the cross correlation matrix between the foreground coefficients of the five mock catalogues and power-spectrum amplitudes from the posterior samples of the corresponding Markov chain. Results are presented in Fig. \ref{fig:corrmat_spec_fg}. It can be seen that correlations and anti-correlations can amount of up to ten percent across all modes in Fourier space.

\begin{figure*}
\centering
  \includegraphics[width=1.\linewidth]{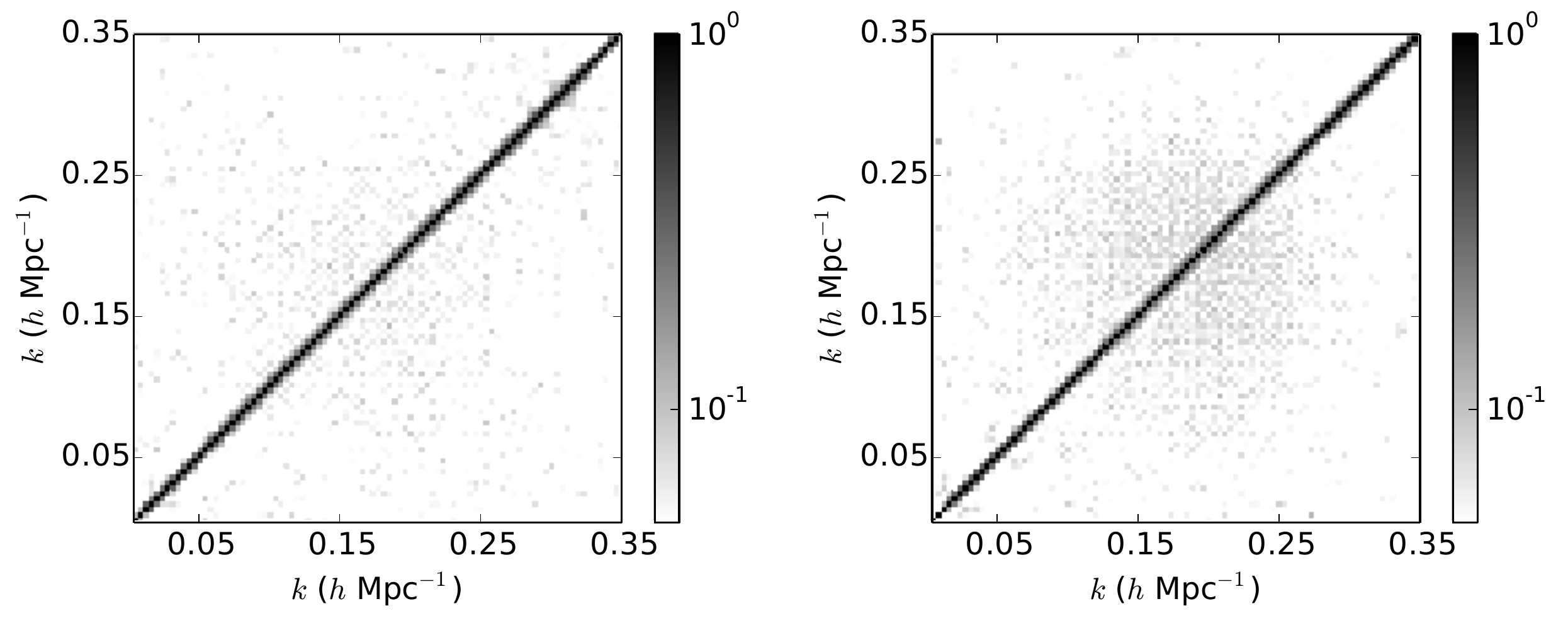}
  \caption{We provide in the two above panels the correlation matrix of power-spectrum amplitudes with respect to their mean value in the case of unaccounted foregrounds (left panel) and modelled foregrounds (right panel). The correlation matrix is normalized using the variance of the power spectrum amplitudes. We note that the colour-map is truncated at a correlation level of $5\,10^{-2}$. We estimate that all values below this threshold are too noisy to be cleanly represented and discard them. \label{fig:corrmat}  }
\end{figure*}

Additionally we tested whether the sampler correctly accounts for the combined effects of foreground contaminations, galaxy biases and unknown noise amplitudes 
by estimating the co-variance matrix of inferred power-spectra from the \ares{} runs. As can be seen in Fig. \ref{fig:corrmat} the co-variance matrix for both runs exhibit strong diagonal shape indicating that the algorithm correctly accounted for the otherwise erroneous mode coupling introduced by survey geometry and foreground effects. Residual off-diagonal contributions amount to less than ten percent. 

These results clearly demonstrate the feasibility of dealing with strong but unknown foreground contaminations when inferring cosmological power-spectra from deep galaxy observations.

\section{Summary and Conclusion}
\label{sec:conclusion}
Major challenges for the analysis of next generation deep galaxy redshift surveys arise from the requirement to account for an increasing amount of systematic and stochastic uncertainties. In particular foreground effects and target contaminations due to e.g. stars and dust, can greatly affect the observation of galaxies. If not accounted for properly, these effects can yield erroneous modulations of galaxy number counts across the sky, which hinders the immediate inference of power-spectra and three dimensional density fields from such galaxy samples.

To address this issue, in this work we have described a fully Bayesian treatment of unknown foreground contamination in the process of inferring cosmological power-spectra from deep galaxy surveys. In particularly we build upon the previously presented Bayesian inference framework \ares{} \citep[for details see][]{JASCHESPEC2010,JaschePspec2013,2015MNRAS.447.1204J}.
The \ares{} algorithm aims to jointly infer three dimensional density fields, corresponding cosmological power-spectrum, luminosity dependent galaxy biases and unknown noise levels. Being a full Bayesian inference engine \ares{} further correctly provides joint and correlated uncertainties for all target quantities by performing an efficient Markov chain Monte Carlo sampling within a block sampling scheme as indicated in Fig.~\ref{fig:flowchart}.

In this work we extend this hierarchical Bayesian framework by also including an additional sampling procedure to account for foreground and target contamination effects. As discussed in Section~\ref{sec:contamination_model} such contaminating effects particularly affect the estimation of the spectroscopic completeness of a given galaxy survey. Naive estimation of the probability of acquisition of galaxy spectra at given positions in the sky, by calculating ratios of observed galaxy spectra and all observed photometric targets, ignores the possibility that photometric targets are contaminated by foreground stars or dust extinction. Such effects are likely to artificially increase or decrease the estimated number of galaxy targets in observations. In consequence these effects introduce an artificial modulation of observed galaxy number densities across the sky, which in turn yields erroneous large scale power in inferred cosmological power-spectra. 

As demonstrated in Section~\ref{sec:contamination_model}, foreground and target contaminations can be accounted for by describing the mismatch between actual galaxies and observed photometric targets as multiplicative correction factors to estimated spectroscopic completeness maps. These correction factors can be described as combinations of various templates for foreground effects, such as introduced by stars or dust, and corresponding unknown template coefficients. The aim of this work is to jointly infer these template coefficients together with the three dimensional density field, corresponding cosmological power-spectra, galaxy biases and unknown noise levels. This goal can be achieved by simply adding an additional sampling scheme for foreground coefficients to the block sampling framework of the \ares{} algorithm. Interestingly, as discussed in section \ref{sec:fg_sampler} and Appendix \ref{app:fg_sampler}, we were able to derive a direct sampling scheme by introducing auxiliary random fields and marginalizing over them. The corresponding algorithm to generate random realizations of foreground template coefficients is given in Algorithm \ref{cond_fg_sampler}. 

We test the performance of improved \ares{} algorithms via applications to mock galaxy observations. To further evaluate the impact of foreground and target contamination effects on the inference of cosmological power-spectra we compare two test scenarios, with and without treatment of foreground effects. Corresponding artificial galaxy observations were self-consistently generated according to the data model described in Section~\ref{sec:mockdata}. In particular, these artificial observations seek to emulate realistic features of the SDSS DR7 main, the LOW-Z and the CMASS galaxy sample. Effectively this results in five artificial galaxy surveys that are jointly handled by the \ares{} framework, while self-consistently accounting for their respective systematic and stochastic uncertainties, including survey geometries, selection effects, galaxy biasing and foregrounds. 

This artificial data was then used to test the statistical performance of our sampling framework. In particular we tested the burn-in behaviour of the algorithm by starting the Markov chain from an over-dispersed state. As described in Section~\ref{sec:stat_eff}, we start the chain with an initial power-spectrum scaled by a factor 0.1. The following burn-in behaviour then manifested itself by a coherent drift of sequential power-spectrum samples towards preferred regions in parameter spaces. We estimated the burn-in phase to be completed after $\sim 2000$ sampling transitions. The statistical efficiency of the sampler was estimated by measuring the correlation length between subsequent posterior power-spectrum samples. As demonstrated in section \ref{sec:stat_eff} the sampler exhibits a correlation length of a few hundred samples. This leaves us with a numerically efficient sampling framework to explore cosmological power-spectra in deep galaxy surveys.
It should be remarked that there exist various possibilities to further improve statistical efficiencies but details are left to future publications.

Results for inferring foreground and target contamination template coefficients are described in section \ref{sec:temp_coeffs}. In particular we have used two realistic foreground templates describing foreground stars in the galaxy and dust extinction. Furthermore, our implementation is general enough to account for an arbitrary amount of foreground templates. To demonstrate the feasibility of inferring these parameters jointly with the cosmological power-spectrum, three dimensional density fields, galaxy biases and noise levels, we presented two dimensional marginalized distributions for template coefficients. These results show that different contamination contributions can be recovered within {$\le$~2.3} sigma of their input values. Given inferred foreground coefficients it is further possible to reconstruct an effective completeness mask and corresponding uncertainties. As expected, in the example tested in this work, uncertainties in recovered completeness masks are largest in regions where stellar contaminations of the target distribution are also large. 

In Section~\ref{sec:density_field}, we have studied the impact of foreground and target contaminations on the inference of three dimensional density fields from galaxy surveys. To do that we have contrasted a run with and without foreground treatment. Ignoring foreground effects when inferring density fields yields excessive large scale power particularly in regions most affected by such contaminations. In contrast detailed Bayesian treatment of foreground systematics yields inferred density fields showing a homogeneous distribution of power across the inference domain. It must be remarked that if foreground contaminations are not explicitly modelled within the data model, then their effects will be attributed to a real signal.  

This result is also in agreement with inferred cosmological posterior power-spectra, as presented in section \ref{sec:power_spectra}. The inferred power-spectrum of the run without foreground treatment agrees with the visual impression obtained from corresponding three dimensional density fields. In particular it reflects the observed excess in large scale power. Ignoring foreground effects may further lead to an overall bias across the entire Fourier domain with respect to the true underlying power-spectrum. In contrast detailed Bayesian foreground treatment yields inferred power-spectra in agreement with respect to the underlying fiducial truth over the entire range of Fourier modes, as considered in this work. 

We have further tested the impact of different foreground effects on the inference of power-spectrum amplitudes by estimating their correlations with foreground template coefficients throughout the full Fourier range. These results show that correlations and anti-correlations can amount up to ten percent throughout large ranges in Fourier space. 

The \ares{} algorithm also accounts for artificial mode coupling between power-spectrum amplitudes as introduced by survey geometries and completeness masks. To demonstrate that fact we have estimated the correlation matrix of power-spectrum amplitudes from our Markov runs. These tests show that residual artificial mode coupling is typically much less than ten percent. These results indicate the validity of the algorithm in scenarios with heavily masked data. We are currently using our method to process the actual BOSS data, which will presented in our companion paper \citep[][]{Guilhem_in_prep}.

As demonstrated in this work detailed treatment of foreground and target contamination is essential to recover unbiased estimates of three dimensional density fields and corresponding cosmological power-spectra from present and next generation surveys. 
The proposed \ares{} algorithm provides a joint and statistically rigorous Bayesian inference framework to achieve this goal and prevent misinterpretation of observations.

\begin{acknowledgements}
This research was supported by the DFG cluster of excellence "Origin and Structure of the Universe" (www.universe-cluster.de).
This work made in the ILP LABEX (under reference ANR-10-LABX-63) was supported by French state funds
managed by the ANR within the Investissements d'Avenir programme under reference ANR-11-IDEX-0004-02.
The Parkes telescope is part of the Australia Telescope which is funded by the Commonwealth of Australia for operation as a National Facility managed by CSIRO.
We acknowledge financial support from "Programme National de Cosmologie and Galaxies" (PNCG) of CNRS/INSU, France. This work was supported in part by the ANR grant ANR-16-CE23-0002.
\end{acknowledgements}

\bibliography{paper}

\onecolumn
\begin{appendix}

\section{Sampling foreground coefficients}
\label{app:fg_sampler}
In this appendix, we derive the sampling procedure for a single foreground template coefficient $\alpha_k$.
As described in Section \ref{sec:fg_sampler}, the full joint distribution of all template coefficients may be
sampled by performing a sequential iterative block sampling procedure.  In the absence of any a priori
information on the foreground coefficient $\alpha_k$ we follow a maximally conservative approach by assuming a
uniform prior distribution. 
Using Equation~\eqref{eq:foreground_LH_single}), the logarithm of the conditional posterior distribution for a foreground coefficient in a single galaxy catalogue can then be written as:
\begin{equation}
\log\mathcal{P}\left(\alpha_k| \{\alpha_n\}\setminus \alpha_k, \{N_i\}, \{\delta_i\}, \{\bar{N}\}\right) = -\frac{1}{2}\sum_{i=0}^{v} \frac{\left[N_i-\bar{N}\,M_{i}\left(\{\alpha_n\} \right)\,R_i\,(1+D_i\,b\,\delta_i)\right]^2}{\bar{N}\, M_{i}\left(\{\alpha_n\} \right)\,R_i\,} 
-\frac{1}{2}\log\left(\bar{N}\,M_{i}\left(\{\alpha_n\} \right)\, R_i\,\right)\, . 
\end{equation}
We can simplify notation by noting that different foreground templates contribute multiplicatively an effective survey response operator. We can therefore collapse all multiplicative foreground contributions, except the one currently under consideration, into the effective survey response operator given as:
\begin{equation}
\tilde{R}_i = \bar{N} R_i \prod_{n \setminus k} (1 - \alpha_n F_{n,i}) \, ,
\end{equation}
where $k$ labels the currently considered foreground template and we have used Equation~\eqref{eq:foreground_mod} to factorize foreground contributions. We further introduce the vector:
\begin{equation}
A_i= \tilde{R}_i (1+D_i\,b\,\delta_i) \, .
\end{equation}
Given this notation the conditional posterior distribution for $\alpha_k$ simplifies to:
\begin{equation}
\log\mathcal{P}\left(\alpha_k| \{\alpha_n\}\setminus \alpha_k, \{N_i\}, \{\delta_i\}, \{\bar{N}\} \right) = -\frac{1}{2}\sum_{i=0}^{v} \frac{\left[N_i-(1 - \alpha_k F_{k,i})\, A_i\right]^2}{(1 - \alpha_k F_{k,i})\,\tilde{R}_i\,} 
-\frac{1}{2}\log\left[(1 - \alpha_k F_{k,i})\, \tilde{R}_i\right]\, . \label{eq:proba_alpha_k}
\end{equation}
The expression can be further compressed by introducing the following indexed quantities:
\begin{align}
  B_i &= N_i - A_i , \\
  C_{k,i} & = A_i\, F_{k,i} \text{, and} \\ 
  \gamma_{k,i} & = \, F_{k,i}\, \tilde{R}_i \, .
\end{align}
Consequently the conditional posterior distribution can be expressed as:
\begin{equation}
\log\left( \mathcal{P}\left(\alpha_k| \{\alpha_n\}\setminus \alpha_k, \{N_i\}, \{\delta_i\}, \{\bar{N}\},\right) \right) = -\frac{1}{2}\sum_{i=0}^{v} \frac{\left(B_i+ \alpha_k C_{k,i}\right)^2}{\left(\tilde{R}_i - \alpha_k \gamma_{k,i}\right)\,} 
-\frac{1}{2}\log\left(\tilde{R}_i - \alpha_k \gamma_{k,i}\right)\, .
\label{eq:conditional_alpha_post}
\end{equation}
In order for Equation~\eqref{eq:conditional_alpha_post} to represent a proper probability distribution the following positivity requirement for the variances needs to hold:
\begin{equation}
\forall i,\, \tilde{R}_i - \alpha_k \gamma_{k,i} > 0
\end{equation}
Similarly it is required that:
\begin{equation}
 \forall i, \,\tilde{R}_i > 0,
\end{equation}
which states that the survey response operator should be positive definite.
To ensure these requirements we split the effective survey response operator $\tilde{R}_i $ as follows: 
\begin{equation}
  \tilde{R}_i = \tilde{R}'_{i,k} + \omega \gamma_{k,i} \, .
\end{equation}
By also requiring $\tilde{R}'_{i,k} > 0$ we yield the following requirement for the scalar quantity $\omega$:
\begin{align}
  \tilde{R}'_{i,k} & = \tilde{R}_i - \omega \gamma_{k,i} > 0 \text{, which brings to} \\
  \omega & < \frac{\tilde{R}_i}{\gamma_{k,i}} \, \forall i \, .
\end{align}
We therefore choose $\omega$ to be:
\begin{equation}
\omega = \min_{i} \left(\frac{\tilde{R}_i}{\gamma_{k,i}}\right) \, .
\end{equation}
Given these definitions one can express the conditional posterior distribution as:
\begin{equation}
\log\left( \mathcal{P}\left(\alpha_k| \{\alpha_n\}\setminus \alpha_k, \{N_i\}, \{\delta_i\}, \{\bar{N}\},\right) \right) = -\frac{1}{2}\sum_{i=0}^{v} \frac{\left(B_i+ \alpha_k C_{k,i}\right)^2}{\tilde{R}'_{i,k} + \left(\omega- \alpha_k\right) \gamma_{k,i}\,} 
-\frac{1}{2}\log\left(\tilde{R}'_{i,k} + \left(\omega- \alpha_k\right) \gamma_{k,i}\right)\, .
\end{equation}
Note that this distribution can be described conveniently as the marginalization over a set of auxiliary fields $\{ t_i\}$:
\begin{equation}
 \mathcal{P}\left(\alpha_k | \{\alpha_n\}\setminus \alpha_k, \{N_i\}, \{\delta_i\}, \{\bar{N}\},\right)  \propto  \prod_i\,\int_{-\infty}^{\infty} \mathrm{d}t_i \frac{\mathrm{e}^{ -\frac{1}{2}\sum_{i=0}^{v} \frac{\left(B_i+ \alpha_k C_{k,i}-t_i\right)^2}{ \left(\omega- \alpha_k\right) \gamma_{k,i}\,}}}{\sqrt{2\pi \,\left(\omega- \alpha_k\right) \gamma_{k,i}}} \, \frac{\mathrm{e}^{-\frac{1}{2}\,\frac{t_i^2}{\tilde{R}'_{i,k}}}}{\sqrt{2\pi\,\tilde{R}'_{i,k}}}\, .
\end{equation}
This approach therefore follows a similar line of reasoning as discussed in our previous work when presenting a messenger field Gibbs sampler \citep{2015MNRAS.447.1204J}. Here we propose to jointly sample the template coefficient parameter $\alpha_k$ with the auxiliary messenger field $t_i$ via a two step block sampling procedure. First we generate realizations of the messenger field $t_i$ conditional on the current value of $\alpha_k$ and then we draw a new value of $\alpha_k$ conditional on the given realization of $t_i$. We loop over this small block ten times to ensure a minimal mixing of the variables.
To simplify notation we introduce the following change of variable:
\begin{equation}
\xi = \omega- \alpha_k \, . \label{eq:xi_omega_transform}
\end{equation}
This yields the joint distribution:
\begin{equation}
 \mathcal{P}\left(\xi , \{t_i\}| \{\alpha_n\}\setminus \alpha_k, \{N_i\}, \{\delta_i\}, \{\bar{N}\},\right)  \propto  \prod_i \frac{\mathrm{e}^{ -\frac{1}{2}\sum_{i=0}^{v} \frac{\left(B_i+ \left(\xi-\omega\right) C_{ki}-t_i\right)^2}{ \xi \gamma_{k,i}\,}}}{\sqrt{2\pi \,\xi \gamma_{k,i}}} \, \frac{\mathrm{e}^{-\frac{1}{2}\,\frac{t_i^2}{\tilde{R}'_{ik}}}}{\sqrt{2\pi\,\tilde{R}'_{i,k}}}\, ,
 \end{equation}
 
\begin{algorithm*}[t]
  \begin{algorithmic}[1]
    \Function{sample\_one\_$\alpha$ }{$\alpha_k,\tilde{R}',B,C, \gamma, \omega, k $}
    	\State $\xi \gets \omega - \alpha_k  $
    	\State $w_k \gets 0$
    	\State $z_k \gets 0$
    	\For{$i = 0 \to (\Call{Length}{B} -1)$}
        \Comment{ we loop over all grid elements of B}
	    	\State $t_i \sim  \mathrm{G}(\mu_{i,k}, \sigma^2_{i,k} ) $ 
        	\Comment{\parbox[t]{.5\linewidth}{
            	Here we generate a new value for the messenger field $t_i$. \newline
             	It is a simple normally distributed value centred on $\mu_{i,k}$ with a variance 
                $\sigma^2_{i,k}$.}
            }
          	\State $w_k \gets w_k + (B_i - C_{k,i} \omega - t_i)^2 / \gamma_{k,i} $
            \Comment{Accumulate $w_k$ and $z_k$.}
          	\State $z_k \gets z_k + (C_{k,i})^2 / \gamma_{k,i} $
    	\EndFor\label{forloop_messenger1}
    	\State $\xi \sim \mathrm{GIG}(w_k, z_k)  $
        \Comment{We sample $\xi$ from a GIG distribution.}
    	\State $\alpha_k \gets \omega-\xi  $
    	\Comment{at that point we have generated a new $\alpha_k$ from the conditional probability distribution in Equation~\eqref{eq:proba_alpha_k}.}
    	\State \Return $\alpha_k$
    \EndFunction
  \end{algorithmic}
  \caption{Algorithm derived in Appendix~\ref{app:fg_sampler} to sample the values of $\alpha_k$, the multiplicative coefficient attached the $k$-th foreground. }\label{cond_fg_sampler}
\end{algorithm*}

As can be easily confirmed, sampling the messenger field $t_i$ amounts to simply generating normal random variates with following means and variances:
\begin{equation}
  \mu_{i,k} = \frac{\tilde{R}'_{i,k}}{\xi \gamma_{k,i}+\tilde{R}'_{i,k}}\,\left(B_i+ \left(\xi -\omega\right) C_{k,i}\right) \, ,
\end{equation}
and
\begin{equation}
  \sigma_{i,k} = \frac{\tilde{R}'_{i,k}\, \xi \gamma_{k,i}}{\xi \gamma_{k,i}+\tilde{R}'_{i,k}} \, .
\end{equation}
To generate realizations of the $\xi$ values, we introduce the following quantities:
\begin{equation}
  w_k=\sum_i \frac{\left(B_i-C_{k,i}\,\omega-t_i \right)^2}{\gamma_{k,i}} = \sum_i  w_{k,i}(t_i)
\end{equation}
and
\begin{equation}
  z_k=\sum_i \frac{\left(C_{k,i}\right)^2}{\gamma_{ki}} = \sum_i  z_{k,i}(t_i) \, .
\end{equation}
With this definition the conditional distribution to sample $\xi$ turns into a generalized inverse Gaussian (GIG) distribution given as:
\begin{equation}
\mathcal{P}\left(\xi | \{t_i\}, \{\alpha_n\}\setminus \alpha_k, \{N_i\}, \{\delta_i\}, \{\bar{N}\},\right) \propto \frac{1}{\left(\xi\right)^{\frac{N_v}{2}}}e^{-\frac{1}{2}\left( \frac{w_k}{\xi} + \xi z_k \right)}\, ,
\end{equation}
where $N_v$ is the number of observed grid elements.
The GIG distribution can be conveniently sampled with standard approaches as described in the literature \citep[see e.g.][]{DAGPUNAR1988}.
Finally to obtain a sample for the foreground coefficient $\alpha_k$ we invert the transformation in Equation~\eqref{eq:xi_omega_transform}:
\begin{equation}
\alpha_k= \omega-\xi \, .
\end{equation} 
The respective realizations of the $t_i$ field are not longer required and are immediately discarded, which amounts to a marginalization over the $t_i$ values. An efficient sampling algorithm that avoids storing the full $t_i$ vector is proposed in Algorithm \ref{cond_fg_sampler}. 

\end{appendix}
\label{lastpage}

\end{document}